\documentclass[seceq]{ptptex}

\usepackage{graphicx}
\usepackage{amsmath,amsthm,amsfonts}
\usepackage{tabularx}

\newcommand{\pr}{Phys. Rev. }

\newcommand{\Apj}{Astrophys. J. }

\title{%        %You can use \\ for explicit line-break
Affleck-Dine Baryogenesis and Heavy Element Production from Inhomogeneous Big Bang Nucleosynthesis
}

\author{%       %Use \scshape  for the family name
Shunji \textsc{Matsuura}$^{1,}$\footnote{E-mail: smatsuura@utap.phys.s.u-tokyo.ac.jp}
Alexander D.\textsc{Dolgov}$^{2,3,4}$
Shigehiro \textsc{Nagataki}$^{5}$
and Katsuhiko \textsc{Sato}$^{1,6}$
}

\inst{%         %Affiliation, neglected when [addenda] or [errata]
$^1$Department of Physics, School of Science, University of Tokyo, 7-3-1 
Hongo, Bunkyo, Tokyo 113-0033, Japan
\\
$^2$INFN, sezione di Ferrara,Via Paradiso, 12 - 44100 Ferrara,Italy
\\
$^3$ITEP, Bol. Cheremushkinskaya 25, Moscow 113259, Russia
\\
$^4$ICTP, Trieste, 34014, Italy
\\
$^5$Yukawa Institute for Theoretical Physics, Kyoto University, Kyoto 
606-8502, Japan
\\
$^6$Research Center for the Early Universe, the University of Tokyo, Tokyo  113-0033, Japan
}

\abst{%         %this abstract is neglected when [addenda] or [errata]
Recent observations have revealed that small-scale structure exists at high 
redshift.
We study the possibility of the formation of such structure during baryogenesis and
big bang nucleosynthesis.
It is known that under certain conditions, 
high density baryonic bubbles are created in the Affleck-Dine model 
of baryogenesis, and these bubbles may occupy a
relatively small fraction of space, while the dominant part of the 
cosmological volume is characterized by the normal observed baryon-to-photon
 ratio, $\eta = 6\cdot 10^{-10}$.
 The value of $\eta$ in the bubbles could be much
larger than the usually accepted value (indeed, even close to unity)
and still be consistent with the existing data on light element abundances and the 
observed angular spectrum of cosmic microwave background radiation (CMBR).
We find upper bounds on $\eta$ by comparing the abundances of
heavy elements produced in big bang nucleosynthesis (BBN) and those of metal poor stars. We 
conclude that $\eta$ should be smaller than $10^{-5}$ in some metal poor star 
regions.
}

\begin{document}
\maketitle
\bigskip

          %%%%%%%%%%%%%%%%%%%%%%%%%%%%%%%%%%%%%%%%
          %               INTRODUCTION           %
          %%%%%%%%%%%%%%%%%%%%%%%%%%%%%%%%%%%%%%%%

\section{Introduction}

One of the biggest puzzles in astrophysics is the nature of structure formation from early stages.
We know that there exist galaxies, quasars (QSO) and stars in our universe.
These structures must have formed at some stage in the evolution of the
universe. But when did the universe become light?
 Recent observations suggest that such structure formations 
began earlier than conventionally thought.
For example, Wilkinson Microwave Anisotropy Probe (WMAP) data suggest that
 reionization began when $z \sim$20 \cite{WMAP}.
Also according to Refs.~\cite{Barth:2003}~\cite{Dietrich:2002}, 
 star formation activity 
started when $z \sim$  10, or at some slightly large value. In addition, 
it is known that the quasar metallicity did not significantly  
change from the time of high redshift to 
the present time~\cite{Boksenberg:2003}.
Some quasars already reached solar or higher metallicity 
when $z$ was no smaller than 6.
Recently a galaxy at $z$=10.0 
was observed~\cite{Pello}.
For IGM, the bulk of metal ejection 
occurred when $z \geq 3$ \cite{Pichon:2003}.
With regard to the abundances of metal poor stars
, see Ref.~\cite{Cohen}. 

Do these QSO, intergalactic medium (IGM) and stars from high redshift contradict the theory of 
structure formation? 

We still do not have a "standard theory" of stars and QSO formation. But 
in some models (for example, \cite{haiman}~\cite{teg}), these observations do not
seem to yield strong contradictions in the cases of QSO and reionization, while the galaxy
observed at z=10 corresponds to the collapse of $\geq$ 2$\sigma$ fluctuations \cite{Pello}.

We can adjust the theory by changing the initial mass function and 
other parameters in order to account for the observations. 
However, because we understood the mechanism of star formation
only poorly and also because observations suggest some
possibility of the existence of structures at higher redshift, it is valuable to consider
an alternative scenario for objects in the early universe that have been or will be
 observed.

We propose a scenario in which the seeds of these structures are produced
during baryogenesis. One of the present authors found~\cite{Dolgov:1993si}
 that under certain conditions, 
 high density baryonic bubbles with small spacial size are produced, while 
most of
the universe is characterized by the small baryon-to-photon ratio ($\eta$) observed through 
BBN~\cite{light1} and CMBR~\cite{cmb},
 $\eta = 6\cdot 10^{-10}$.
 The model 
represents a modified version of the Affleck-Dine~\cite{affl-dine}
baryogenesis scenario and is based on the hypothesis that
the Affleck-Dine field $\phi$ is coupled to the inflaton field $\Phi$.
The interaction Lagrangian is assumed to have the general renormalizable 
form
\begin{equation}
  \begin{split}
    \mathcal{L}_{int} &= \lambda |\phi|^2 \Phi ^2 + g|\phi|^2 \Phi  \\
                &= \lambda (\Phi - \Phi _1)^2|\phi |^2-
\lambda \Phi ^2_1|\phi |^2 , \end{split}
\end{equation}
where $g$ and $\lambda$ are the coupling constants and $\Phi _1=-g/2\lambda $.
It is known that
the effective mass of the field $\phi$ may contain the contributions
\begin{equation}
(m^{\phi}_{eff})^2=m^2_0 +\xi R+\beta T^2+ \lambda (\Phi - \Phi _1)^2  ,
\end{equation}
where $\xi$ and $\beta$ are constants, $R$ is the curvature scalar and $T$ is 
the temperature of the primeval plasma, 
$m^2_0 $ is the vacuum mass of $\phi$, 
barring the contribution from $(\lambda_1 \Phi _1)^2$. 
For minimal coupling of $\phi$ to the gravity field, the coupling to the curvature vanishes,
by definition. However, radiative corrections may induce relatively
small coupling, with $\xi \leq 10^{-2}$. Temperature corrections to the
mass appear at higher orders in the perturbation theory, and thus usually
$\beta \ll 1$. In what follows, we ignore these two contributions,
as they are not essential for the dynamics of the formation of high B bubbles
(but may be important for some details of their evolution). We assume,
though it is not necessary, that $m_0^2 \sim H_I^2$, where $H_I$ is
the Hubble parameter during inflation. The coupling constant $\lambda$
is bounded from above by the condition that the inflation self-interaction
be sufficiently weak. The interaction (1.1) induces the inflaton
self-coupling $\sim \lambda^2 \Phi^4$, and the condition 
$\lambda \sim 10^{-5}$--$10^{-6}$ allows it to remain on a safe level to
avoid density perturbations that are too large. An essential condition for the
realization of the high-B-bubble scenario is a negative effective mass
squared of the field $\phi$ when the inflaton field, $\Phi$, is close to 
$\Phi_1$. To this end, we stipulate the condition
\begin{equation}
m^2 _0 + \xi R < 0,
\end{equation}
and because we assumed that the term $\xi R$ is negligible, this 
condition is satisfied for negative $m_0^2$. In fact, a weaker
condition is possible, namely that $m_0^2 < H_I^2(\Phi_1)$, 
where $H_I (\Phi_1)$ is the Hubble parameter for $\Phi \sim \Phi_1$,
and it is assumed that inflation still proceeds at this stage. 
This condition is
necessary, in particular, to ensure a sufficiently large size of the bubbles.

As argued in Ref.\cite{Dolgov:1993si} , the scenario of the formation of
bubbles with high baryon asymmetry would be relevant, for example, for
$H_I \sim (10^{-5}$--$10^{-6}) m_{Pl}$, $\lambda \sim 10^{-10}$--$10^{-12}$,
and $\Phi_1 \sim m_{Pl}$, but other wider ranges of parameter values 
also seem to be allowed.

The inflaton field $\Phi$ is assumed to evolve from some high value
$\Phi > m_{Pl}$ down to zero, where inflation ends. In the course of
this process, the effective mass squared of $\phi$ changes from a positive
value to a negative value (or to $|m_{eff}| < H_I$),
and then back to a positive value. 

An important assumption is that the potential of the field $\phi$
possesses two minima at small values of $m_{eff}$. 
This is true, e.g., for the
Coleman-Weinberg\cite{cw} potential
\begin{equation}
U(\phi) = m^2_{eff} |\phi|^2 + 
\lambda_\phi |\phi|^4 \ln \frac{|\phi|^2}{\sigma^2} .
\label{u-cw}
\end{equation}
We assume that chaotic inflation is valid, though not a necessary condition. 
Initially when the inflaton field has a large value, i.e. $\Phi > m_{Pl}$,
and the square of the effective mass is positive, the potential
(\ref{u-cw}) has only one minimum, at $\phi =0$. As $\Phi$ decreases,
the effective mass decreases, and a new minimum at $\phi \sim \sigma$
appears. When $m^2_{eff}$ becomes negative or zero, the minimum at $\phi=0$
becomes a local maximum, and the field $\phi$ tends to increase from 0 to a larger
value in the direction of the minimum at $\phi \sim \sigma$. If $m_{eff}^2$
remained negative for a sufficiently long time, the field $\phi$ would tend
to this other minimum in the entire space. However, because $m^2_{eff}$
remains negative for only a finite time, depending upon the duration
of the ``negative'' period and the magnitude of the fluctuations of 
$\phi$ near $\phi = 0$ (the latter being typically of order 
$\delta\phi \sim H$), only some bubbles with large values $\phi\sim\sigma$
occupying a small fraction of space are formed, while the rest of
space is characterized by $\phi =0$. 

As $\Phi$ decreases further, the effective mass becomes positive again,
and the second minimum at $\phi \sim \sigma$ becomes higher than the first
one at $\phi=0$, and ultimately it disappears. Subsequently the field $\phi$
stuck in this minimum decreases to zero. If this field carries
baryonic charge, as in the Affleck-Dine scenario of baryogenesis,
then a large baryon asymmetry is generated inside such bubbles, while
in the region of space occupied by $\phi$ near zero, the asymmetry is
much smaller. This asymmetry may be generated by the decay of the same $\phi$ field
as in the bubbles but with a smaller amplitude or by some other mechanism of 
baryogenesis, which is usually less efficient and creates an asymmetry much smaller 
than that in the Affleck-Dine case. It is worth noting that the scenario of
baryognesis inside the bubbles suggested in Ref.~\cite{Dolgov:1993si}
and used here
is very similar to the original Affleck-Dine scenario with the only difference
being that in the original version, the field $\phi$ is displaced from the minimum
at $\phi=0$ along the flat directions of the potential by quantum
fluctuations during inflation, while here the field $\phi$ is displaced from
zero to a large value by a negative $m_{eff}^2$. Here it is worth repeating that because
$m_{eff}^2$ remains negative during only a short time interval, 
only a small fraction of $\phi$ is able to reach the minimum near the large
$\phi=\sigma$.

As a result of the process described above,
the bulk of the universe would come to possess the normal small
baryon asymmetry observed through BBN and CMBR, $\eta = 6\cdot 10^{-10}$, 
while in some small regions, the asymmetry could be much larger, perhaps even
reaching $\eta \sim 1$. Depending upon the 
details of the $\phi$ potential and the mechanism of CP-violation (though
in some versions of the scenario, explicit CP-violation is unnecessary),
the value of the asymmetry may vary form bubble to bubble or remain
the same, in which case only the bubble size may vary. The baryon asymmetry 
inside the bubbles cannot be reliably predicted because of the existence of many unknowns
in the theory, but typically, it should be the same as in the usual Affleck-Dine
model.

After baryogenesis, the initial energy density difference between the
interior of the bubbles and external space is small (isocurvature
perturbations). When the QCD phase transition takes 
place and quarks form nonrelativistic baryons, large density 
inhomogeneities develop, because the equation of state inside the 
bubbles begin to deviate from the relativistic one, $p=\rho /3$, 
which is valid in the rest of the universe. The subsequent 
destiny of these high 
density baryonic bubbles depends on the size of the bubbles and the value 
of the baryonic charge asymmetry. Some of them may form unusual stars or 
anti-stars with a high initial fraction of heavy nuclei, and others may form
primordial black holes.

%Most of them would become primordial black holes.
 
According to the calculations of Ref.~\cite{Dolgov:1993si} ,
the mass distribution in these region is
\begin{equation}
\frac{dN}{dM} \propto \exp \left( -C \ln^2 \frac{M}{M_1} \right), 
\end{equation}
where C and $M_1$ are constants which should be determined by
$H_I$, the time evolution of $m_{eff}$ when $\Phi \simeq \Phi _1$, 
the time width $\Phi = \Phi _1$ and $\Phi = 0$, etc., which in turn are
determined by the unknown details and parameters of the potentials of 
the inflaton and $\phi$ fields.
``Natural'' values of these constants are $C \leq 1$ and 
$M_1 = (10^{-3}$--$10^6)M_\odot$. 
The distribution of bubble sizes is similar. We assume that the
bubble sizes are smaller than the galaxy size. 

Because this model has a great degree of freedom and the parameters can be 
treated as free, we cannot a priori restrict such important quantities
as, e.g., the bubble size and the baryon density inside the bubbles.
Instead, assuming that this model is applicable to the early universe
and studying its implications for the formation of primordial objects, dark 
matter, elemental abundances, etc., one could either find observational
confirmation of the discussed mechanism in the generation of large baryonic
inhomogeneities at small scales or obtain bounds on the magnitude of the
effects and thus on the parameter values used in the model.

In order to carry out either of the above stated tasks,
 we need observational data concerning BBN and CMBR.
Observations of
the abundances of primordially produced light elements~\cite{light1} 
 allow the
``measurement'' of the baryon-to-photon ratio $\eta = n_B/n_\gamma$ during BBN.
On the other hand, the spectrum of angular fluctuations 
of CMBR~\cite{cmb} 
also yields a value of
$\eta$ that is in good agreement with BBN result~\cite{bbncmb}. 
Based on the observational
 BBN and CMBR data, many interesting restrictions on 
models of baryogenesis, the magnitude of the reheating temperature, unobserved particle species,
 their masses, lifetimes, and coupling strengths,
 possible types of phase transitions in the early universe, etc., have been obtained.
Most of these studies deal with standard (or homogeneous) big bang 
nucleosynthesis (SBBN)~\cite{SBBN}, or inhomogeneous big bang 
nucleosynthesis (IBBN)~\cite{IBBN}, 
but with a very small magnitude of the fluctuations of the baryonic charge 
density.
However our case is different. In our model, 
large inhomogeneities in spatially small regions are essential, and from the 
observation of small scale structure, we
determine restrictions on the nature of elementary physics.
 Though  CMBR supports the existence of scale invariance in the primordial power spectrum, 
recent observations suggest it is reasonable that there was
small scale and small fraction of structure in the early universe.
Motivated by these findings, we consider here the hypothesis that 
the presently existing metals could have been produced in the very early universe during
BBN. We have computed the abundances of different (not only light)
elements produced in the bubbles with high baryon density and compared the 
abundances of heavy elements produced in these bubbles during BBN
with those in QSO, IGM and metal poor stars observed at high redshifts.

%\section{The calculation of big bang nucleosynthesis}
\section{Heavy element production at BBN with a large baryon number}

We assume that the characteristic bubble size is much larger than the
baryon diffusion length and hence that baryon diffusion is not important.
In this case, the problem is greatly simplified, because we need only 
consider homogeneous nucleosynthesis. The reaction network used in
this work was applied previously to supernova nucleosynthesis 
calculations \cite{nagataki:1998} \cite{nagataki:1998sa} , but we add 16 
light nuclei. The network includes the 258 isotopes listed in Table 1,
$Z=0$ -- $32$ and $A=1$ -- $74$.

%\begin{table*}[!ht]
%\caption{\label{starlist}Basic observations parameters of the observed stars.}
\begin{table}[htbp]
\begin{center}
\begin{tabular}{llllllllll} 
N & H1 &  &  &  &  &  &  &  &  \\
 & H2 & He3 &  &  &  &  &  &  &  \\
 & H3 & He4 &  &  &  &  &  &  &  \\
 &  & He5 & Li6 & Be7 & B8 &  &  &  &  \\
 &  &  & Li7 & Be8 & B9 &  &  &  &  \\
Ne18 &  &  & Li8 & Be9 & B10 & C11 & N12 &  &  \\
Ne19 & Na20 &  &  & Be10 & B11 & C12 & N13 & O14 &  \\
Ne20 & Na21 & Mg22 &  &  & B12 & C13 & N14 & O15 &  \\
Ne21 & Na22 & Mg23 & Al24 &  &  & C14 & N15 & O16 & F17 \\
Ne22 & Na23 & Mg24 & Al25 & Si26 &  &  &  & O17 & F18 \\
Ne23 & Na24 & Mg25 & Al26 & Si27 & P28 &  &  &  &  \\
 & Na25 & Mg26 & Al27 & Si28 & P29 &  &  &  &  \\
 & Na26 & Mg27 & Al28 & Si29 & P30 & S31 & Cl32 & Ar35 & K36 \\
 &  &  & Al29 & Si30 & P31 & S32 & Cl33 & Ar36 & K37 \\
Ca39 &  &  & Al30 & Si31 & P32 & S33 & Cl34 & Ar37 & K38 \\
Ca40 & Sc40 &  &  & Si32 & P33 & S34 & Cl35 & Ar38 & K39 \\
Ca41 & Sc41 &  &  & Si33 & P34 & S35 & Cl36 & Ar39 & K40 \\
Ca42 & Sc42 & Ti43 &  &  & P35 & S36 & Cl37 & Ar40 & K41 \\
Ca43 & Sc43 & Ti44 &  &  & P36 & S37 & Cl38 & Ar41 & K42 \\
Ca44 & Sc44 & Ti45 & V46 &  &  &  & Cl39 & Ar42 & K43 \\
Ca45 & Sc45 & Ti46 & V47 & Cr48 &  &  & Cl40 & Ar43 & K44 \\
Ca46 & Sc46 & Ti47 & V48 & Cr49 & Mn50 &  &  & Ar44 & K45 \\
Ca47 & Sc47 & Ti48 & V49 & Cr50 & Mn51 & Fe52 &  & Ar45 & K46 \\
Ca48 & Sc48 & Ti49 & V50 & Cr51 & Mn52 & Fe53 &  &  & K47 \\
Ca49 & Sc49 & Ti50 & V51 & Cr52 & Mn53 & Fe54 & Co54 &  & K48 \\
 & Sc50 & Ti51 & V52 & Cr53 & Mn54 & Fe55 & Co55 &  &  \\
 & Sc51 & Ti52 & V53 & Cr54 & Mn55 & Fe56 & Co56 & Ni56 &  \\
 &  &  & V54 & Cr55 & Mn56 & Fe57 & Co57 & Ni57 & Cu58 \\
Zn60 &  &  &  &  & Mn57 & Fe58 & Co58 & Ni58 & Cu59 \\
Zn61 & Ga63 & Ge64 &  &  & Mn58 & Fe59 & Co59 & Ni59 & Cu60 \\
Zn62 & Ga64 & Ge65 &  &  &  & Fe60 & Co60 & Ni60 & Cu61 \\
Zn63 & Ga65 & Ge66 &  &  &  & Fe61 & Co61 & Ni61 & Cu62 \\
Zn64 & Ga66 & Ge67 &  &  &  &  & Co62 & Ni62 & Cu63 \\
Zn65 & Ga67 & Ge68 &  &  &  &  & Co63 & Ni63 & Cu64 \\
Zn66 & Ga68 & Ge69 &  &  &  &  & Co64 & Ni64 & Cu65 \\
Zn67 & Ga69 & Ge70 &  &  &  &  &  & Ni65 & Cu66 \\
Zn68 & Ga70 & Ge71 &  &  &  &  &  &  & Cu67 \\
Zn69 & Ga71 & Ge72 &  &  &  &  &  &  & Cu68 \\
Zn70 & Ga72 & Ge73 &  &  &  &  &  &  &  \\
Zn71 & Ga73 & Ge74 &  &  &  &  &  &  &  \\
\end{tabular}
\end{center}
\caption{Isotopes that are included in our nuclear reaction network}
\label{}
\end{table}

The initial and final temperatures are $10^{11}$K and $10^{7}$K, 
which correspond to the time interval from $10^{-2}$ to $\sim 10^{6}$ sec.

In our calculation of the time evolution of the baryon density and temperature, 
we use the Friedmann equation
\begin{equation}
H^2=\frac{8\pi}{3}G \rho _{total}  , 
\end{equation}
where $\rho _{total}= \rho _{\gamma} + (\rho _{e^-}+\rho _{e^+}) + 
\rho _{\nu} +\rho _b$, and the energy conservation law
\begin{equation}
\frac{d}{dt}(\rho R^3)+\frac{p}{c^2}\frac{d}{dt}(R^3)+
R^3\frac{d\rho}{dt}|_{T=const}=0 ,
\end{equation}
where the last term takes into account the change of energy introduced by 
nucleosynthesis. We do not consider the possibility of neutrino degeneracy.

Calculations of nuclei production during BBN with high $\eta$ have already 
been carried out by one of the present authors. \cite{sato:1985}
 There are 
basically two distinctions between the present work and the that work. The first is 
that the network used in the previous work includes 72 isotopes, while ours includes 258 nuclei.
Therefore, we can predict which heavy elements should be observed, unlike in the case of the
previous work. 
The second is that the main interest of the previous work was to study the 
effects on BBN of a large cosmological
lepton asymmetry with a magnitude
 greater than that of the baryonic asymmetry, while we assume that the lepton
asymmetry is negligibly small.

\section{Results and Discussion}

In our model, the only free parameter is $\eta$.
We carried out the calculation for values of $\eta$ ranging from $10^{-12}$ to $3 \times 10^{-4}$ 
and. Table 2 displays the mass fractions of the main 
product elements.

\begin{table}[htbp]
\begin{tabular}{cc|cc|cc}
\hline
$\eta = $ &$10^{-4}$ & $\eta =$ & $10^{-6}$ & $\eta =$ & $10^{-10}$ \\
\hline
H1 & $6.36\times 10^{-1}$ & H1 & $6.91\times 10^{-1}$ & H1 & $7.77\times 10^{-1}$ \\
He4 & $3.64\times 10^{-1}$ & He4 & $3.09\times 10^{-1}$ & He4 & $2.22\times 10^{-1}$ \\
Ge74 & $6.22\times 10^{-6}$ & He3 & $6.24\times 10^{-7}$ & H2 & $7.37\times 10^{-4}$ \\
Ti44 & $1.79\times 10^{-6}$ & Be7 & $4.96\times 10^{-7}$ & He3 & $7.45\times 10^{-5}$ \\
Ca40 & $9.31\times 10^{-7}$ & C11 & $4.54\times 10^{-8}$ & H3 & $3.49\times 10^{-6}$ \\
Sc43 & $9.41\times 10^{-8}$ & N13 & $3.25\times 10^{-10}$ & Li7 & $2.82\times 10^{-9}$ \\
O16 & $5.59\times 10^{-8}$ & O16 & $1.05\times 10^{-10}$ &  &  \\
Ge72 & $4.65\times 10^{-8}$ & Li7 & $5.31\times 10^{-11}$ &  &  \\
Ca42 & $4.10\times 10^{-8}$ & C12 & $3.55\times 10^{-11}$ &  &  \\
Ca41 & $3.58\times 10^{-8}$ & N12 & $1.90\times 10^{-11}$ &  &  \\
\hline
\end{tabular}
\caption{Main product nuclei and their mass fractions for 
$\eta=10^{-4}, 10^{-6}$ and $10^{-10}$. This result shows that the amount of
heavy elements produced increases as $\eta$ increases. We also see
that many of the product nuclei are proton rich. }
\end{table}

As $\eta$ becomes larger, heavy elements begin to be produced more efficiently. 
A very interesting feature of our result is that the nuclear 
reactions proceed along the proton rich side. The usual nucleosynthesis of 
heavy elements in supernovae proceeds along 
the neutron rich side (r-process). However our calculations show that
BBN produces proton rich nuclei. Though we cannot conclude that 
BBN is a p-process, we can say that most of the product elements are proton 
rich. To confirm this, we calculated the value of Ye$=n_p/(n_p+n_n)$.
The results are presented in Table 3.

%\hspace{2zw}
\begin{table}[htbp]
\begin{tabular}{cc}
\hline
$\eta$ & Ye \\
\hline
$10^{-5}$ & 0.834 \\
$10^{-6}$ & 0.845 \\
$10^{-7}$ & 0.855 \\
$10^{-8}$ & 0.865 \\
$10^{-9}$ & 0.874 \\
$10^{-10}$ & 0.888 \\
\hline
\end{tabular}
\caption{Ye for $\eta = 10^{-10}$ -- $10^{-5}$. It is clear that the 
nuclear reactions proceed in a very proton rich environment. This is the 
reason for the abundance of proton rich nuclei.}
\end{table}
%\hspace{2zw}

It is apparent from Table 3 that the nuclear reactions proceed in a
proton rich environment. The value of Ye is determined by the amount of He 4 
produced, and hence the $\beta$ decay effect is almost negligible.
The $\eta$ dependence of each element abundance is displayed in Fig. $1$.

\begin{figure}[htbp]
\begin{center}
\rotatebox{270}{
\includegraphics[height=8.5cm,width=6cm,clip]{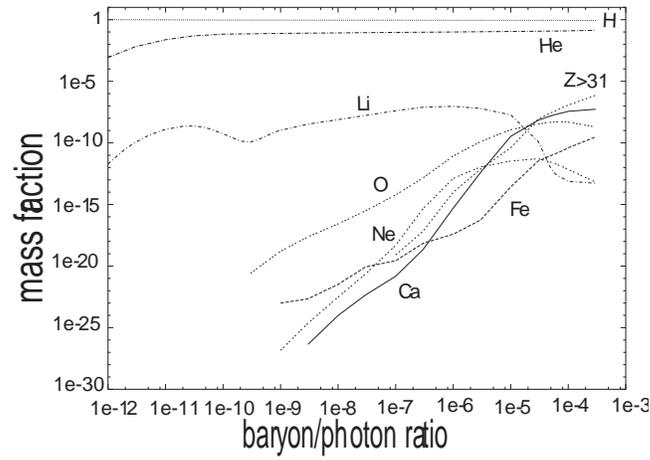}
}

\caption{$\eta$ dependence of the mass fraction of each element.
These of H and He are almost constant. Contrastingly, heavy elements are 
produced in greater abundance as $\eta$ increases, but eventually they begin to be 
consumed to produce even heavier elements.}
\end{center}
\end{figure}

The abundances of H and He remain almost constant over a wide range of values of $\eta$.
C and O increase as $\eta$ increases, but eventually they reach maximum values and 
then decrease. Beyond that point, heavier elements, such as Ca and Fe, begin to dominate.
Two typical examples of the time evolution of the abundance of each element are plotted 
in Figs. $2$ and $3$.

\begin{figure}[htbp]
\begin{center}
\rotatebox{270}{
\includegraphics[height=8.5cm,width=6cm,clip]{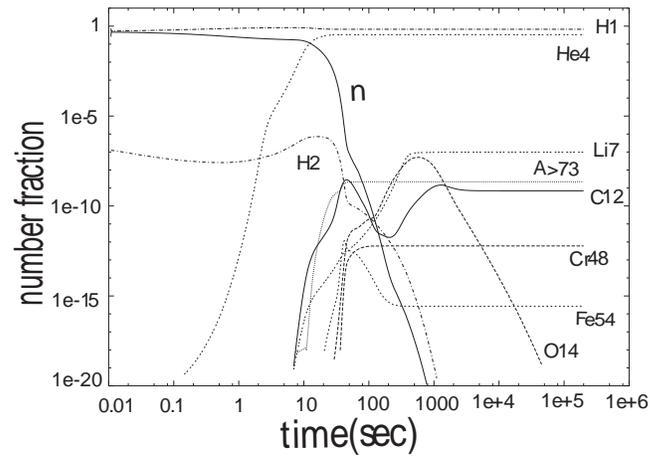}
}
\caption{Time evolution with $\eta =10^{-5}$}
\end{center}
\end{figure}

\begin{figure}[htbp]
\begin{center}
\rotatebox{270}{
\includegraphics[height=8.5cm,width=6cm,clip]{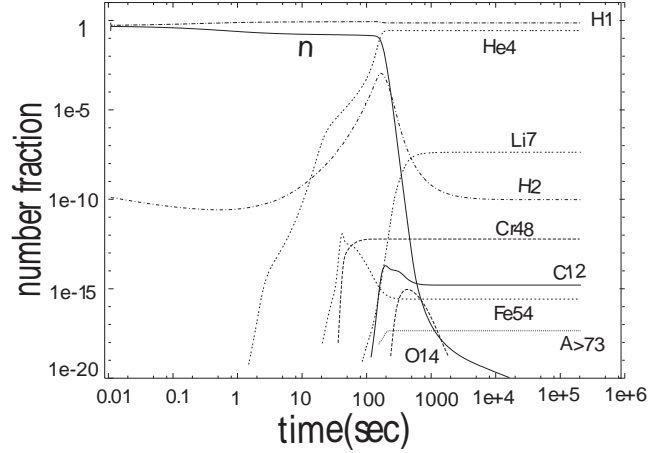}
}
\caption{Time evolution of nuclei abundances with $\eta =10^{-8}$}
\end{center}
\end{figure}

In the early stages, there is an abundance of light elements. Then as reactions 
proceed the light elements are consumed, and heavy elements begin to be 
produced. Production of heavy elements at $\eta = 10^{-5}$ is much larger 
than at $\eta = 10^{-8}$. In particular, nuclei with mass numbers  
greater than 73 begin to dominate at $\eta =10^{-5}$.

In Fig. $4$ the ratio of the BBN number fraction with respect to the present
solar system number fraction is presented. We find two 
peaks in the graph, one corresponding to calcium at Z=20, one corresponding 
to Germanium at Z=32. 
We believe that the peak value at Z=32 should not regarded as the Ge fraction but,
 rather, as 
the sum of the abundances of all elements with atomic number greater 
than 32, because our network includes only those nuclei with  Z$\leq$32, and thus the reaction 
cannot proceed beyond Germanium, and the nuclei accumulate there.
By contrast, the peak at Z=20 should be considered as representing a peak of calculation. 
To understand 
why Ca is produced to much a great degree, it is necessary to know which of the Ca isotopes 
is produced in greatest abundance. These data are presented in Table 4.
\begin{table}[htbp]
\begin{tabular}{cc}
\hline
isotopes & abundances  \\
\hline
Ca40 & $5.33\times 10^{-8}$ \\
Ca41 & $1.8\times 10^{-9}$ \\
Ca42 & $1.93\times 10^{-9}$ \\
Ca43 & $1.21\times 10^{-12}$ \\
Ca44 & $1.89\times 10^{-17}$ \\
\hline
\end{tabular}
\caption{The abundances of Ca isotopes. This shows that Ca 40, which is 
the double magic number nucleus, is dominantly produced.}
\end{table}
The table shows that the dominant output is for Ca 40, which has 20 protons 
and 20 neutrons. It is to be noted that 20 is a magic number,
 and  Ca 40 is a double magic number 
nucleus which is known to be very stable. Surprisingly, no peak can be found 
at Z=26 (Fe). On the contrary, there is a decrease there.
There are two reasons for this. The first is that because of the
p-process, the production of Fe is inhibited. The second is that 
the solar abundance of Fe is large.
At $\eta = 3\times10^{-4}$, [Fe/H]$=-6.70$ is still too small,
but [Ca/H]$=-2.17$ is large enough to be inconsistent with the observations of 
extremely metal poor stars.
For example, [Ca/H]$=-5.37$ for He0107-5240 \cite{Christlieb:2003}.
In order for the Ca abundance produced at BBN to be below this value,
$\eta$ of a high baryon density bubble, if surrounding this metal poor 
star, should be smaller than $10^{-5}$. However, there is such a great degree of
freedom in our model 
that we cannot strongly restrict it with data from only a single observation.
We can only say that if such a  
model is realized, $\eta$ of the bubble around this metal poor star
must be under $10^{-5}$. On the other hand, it may be the case that 
the observed early metal poor
stars are well outside of the baryon rich regions. For confirmation of
the model, we need to carry out
 calculations with larger values of $\eta$,
 because for $\eta \leq 3\times 10^{-4}$,
the abundances of heavy elements are still small
 in comparison with those observed in IGM, around QSO, and  in most of
 the metal poor stars. 
Observations of objects in the early universe
 with anomalously high metal abundances are also desirable.
To impose some restrictions on the mass of the Affleck-Dine 
field, its coupling strength, etc., we need additional observational data. 
Extending our calculations to larger values of $\eta$ in order to obtain predictions for the 
abundances of heavier elements, it may be possible to distinguish usual
stars from our bubble-made stars, which is necessary to improve the restriction. 
With such progress, in the future, we may be able to reach a more definite conclusion.

There is another interesting possibility. To this point, we have considered the case in which 
baryons in the bubbles do not diffuse. However, if the bubbles are smaller than the 
quark diffusion distance, which can be evaluated \cite{Dolgov:1993si} 
in comoving coordinates as $\sim$ $(tl_{free})^{1/2}$, where $l_{free}$ is the quark mean free path,
$l_{free} \sim (\sigma N)^{-1} \sim T^{-1}$, they do diffuse. This 
could affect the angular spectrum of CMBR. Most IBBN studies carried out to this time
treat only 
small values of $\eta$, specifically, $\eta \sim 10^{-10}$. However, with our approach, we are able to 
investigate a novel relation between IBBN and CMBR. Our study also suggests 
some possibilities for the origin of p-nuclei in the solar system. To investigate this, 
we need to calculate the BBN abundances with a reaction network that 
includes heavier nuclei. There is also the possibility that the baryon rich
bubbles, though not forming primordial black holes, might end up as stellar-type
or planetary-type objects. In this case (which is under investigation), the
observational consequences could be quite different.

\begin{figure}[htbp]
\begin{center}
\rotatebox{270}{
\includegraphics[height=8.5cm,width=6cm,clip]{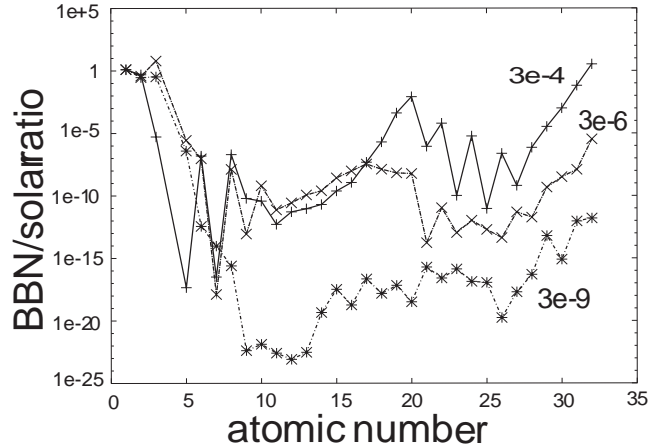}
}
\caption{The ratio of BBN to the solar mass fraction for a range of elements.
There is a prominent peak corresponding to
Ca. We compare this Ca 
abundance with observation results to obtain an upper bound on $\eta$.}
\end{center}
\end{figure}

\section{Acknowledgments}

S.M. thanks Kazuhiro Yahata, Tomoya Takiwaki, Yasuko Hisamatsu and Kumiko Kihara
for helping me in computer and Kazuhiro Yahata, Kohji Yoshikawa and  Atsunori Yonehara
for useful discussions.  
This research was supported in part by Grants-in-Aid for Scientific
Research provided by the Ministry of Education, Science and Culture
of Japan through Research Grant No.S 14102004, No.14079202 and No.16740134.

%%%%%%%%%%%%%%%%%%%%%%%%%%%%%%%%%%%%%%%%%%%%%%%%%%%%%%%%%%%%%%%%%%%%%%%%%%%%
%\bibliography{2-first}

\end{document}